\documentclass[a4paper,fleqn,usenatbib]{mnras}

\usepackage{newtxtext,newtxmath}
\usepackage{pifont}
\newcommand{\xmark}{\ding{55}}
\newcommand{\cmark}{\ding{51}}

\usepackage[T1]{fontenc}
\usepackage{ae,aecompl}

\usepackage{graphicx}	
\usepackage{amsmath}	
\usepackage{verbatim}

\title[galactic anatomy II]{The Anatomy of a Star-Forming Galaxy II: FUV heating via dust}

\author[S. M. Benincasa et al.]{S. M. Benincasa,$^{1,2}$\thanks{E-mail: smbenincasa@ucdavis.edu}
J. W. Wadsley,$^{2}$
H. M. P. Couchman,$^{2}$
A. R. Pettitt,$^{3}$
\newauthor
B. W. Keller$^{4}$,
R. M. Woods$^{2}$
 and J. J. Grond$^{2}$
\\
$^{1}$Department of Physics, University of California, Davis, 1 Shields Ave., Davis, CA 95616, United States\\
$^{2}$Department of Physics and Astronomy, McMaster University, L8S 4M1 Hamilton, Canada\\
$^{3}$Department of Physics, Faculty of Science, Hokkaido University, Sapporo 060-0810, Japan\\
$^{4}$Astronomisches Rechen-Institut, Zentrum f{\"a}r Astronomie der Universit{\"a}t Heidelberg, M{\"o}nchofstra{\ss}e 12-14, D-69120 Heidelberg, Germany 
}

\date{Accepted XXX. Received YYY; in original form ZZZ}

\pubyear{2019}

\begin{document}
\label{firstpage}
\pagerange{\pageref{firstpage}--\pageref{lastpage}}
\maketitle

\begin{abstract}
Far-Ultraviolet (FUV) radiation greatly exceeds ultraviolet, supernovae and winds in the energy budget of young star clusters but is poorly modeled in galaxy simulations.  We present results of the first isolated galaxy disk simulations to include photo-electric heating  of gas via dust grains from FUV radiation self-consistently, using a ray-tracing approach that calculates optical depths along the source-receiver sight-line. This is the first science application of the TREVR radiative transfer algorithm.  We find that FUV radiation alone cannot regulate star formation. However, FUV radiation produces warm neutral gas and is able to produce regulated galaxies with realistic scale heights.  FUV is also a long-range feedback and is more important in the outer disks of galaxies.  We also use the super-bubble feedback model, which depends only on the supernova energy per stellar mass, is more physically realistic than common, parameter-driven alternatives and thus better constrains supernova feedback impacts.
FUV and supernovae together can regulate star formation without producing too much hot ionized medium and with less disruption to the ISM compared to supernovae alone.  
\end{abstract}




\section{Introduction}

Star formation is effectively regulated by how galaxies make the cold, dense clouds where stars form today.  FUV emission from massive stars heats the surrounding medium via the photoelectric effect \citep{watson1972, draine1978, wolfire1995}. This FUV emission is often the dominant heating mechanism for the interstellar medium (ISM) and thus controls the warm vs. cold phase balance. FUV is not absorbed by atomic hydrogen, giving it a long mean free path (of order kpc) and it is thus far-reaching for dust surface densities characteristic of most nearby galaxies.  Though FUV heating is strongest near young star clusters, it permeates the ISM and is non-local compared to other feedback such as ultraviolet radiation, winds and supernovae.  This work seeks to better understand the impact of both FUV and supernovae on star formation and the ISM in a typical nearby galaxy.

In galaxy evolution, gas and star formation are intimately connected.  The Kennicutt-Schmidt relation links the star formation rate surface density with the total gas surface density, $\dot{\Sigma}_* \sim \Sigma_g^N$, where $N\sim1.4$ \citep{schmidt1959, kennicutt1998}.  The star forming main sequence demonstrates a tight correlation between the galactic star formation rate and the total stellar mass of a galaxy  \citep{brinchmann2004, daddi2007, noeske2007}.  These relationships have been established on local scales as well: on the kpc-scale for the star-forming main sequence \citep[e.g.][]{hsieh2017} and the sub-kpc scale for the Kennicutt-Schmidt relation \citep[e.g.][]{bigiel2008}. The original Kennicutt-Schmidt (KS) relation probed typical surface densities $\Sigma_g \gtrsim 10$ M$_{\odot}$/pc$^2$ \citep{kennicutt1998}.  For $\Sigma_g \lesssim 10$ M$_{\odot}$/pc$^2$ the KS relation steepens significantly \citep[][and references therein]{kennicutt2012}.

At high surface densities, $\Sigma_g \gtrsim 100$ M$_{\odot}$/pc$^2$, combinations of supernovae and radiation pressure are commonly invoked to regulate star formation \citep{ostriker2011, shetty2012}.  For surface densities between $10 \lesssim \Sigma_g \lesssim 100$ M$_{\odot}$/pc$^2$, more representative of nearby galaxies, we expect supernova feedback and stellar UV/FUV heating to play roles in regulating the star formation rate. At even lower surface densities, $\Sigma_g \lesssim 10$ M$_{\odot}$/pc$^2$, star formation becomes much less effective.  Here a key requirement for star formation may be the presence of two thermal phases in the ISM \citep{elmegreen1994}.  The absence of a persistent cold phase might sharply limit star formation in outer galactic disks \citep{schaye2004}. 

At all surface densities, there is scatter from a single power-law in the Kennicutt-Schmidt relation. In some regions this scatter spans almost 2 dex in $\dot{\Sigma}_*$.  This intrinsic scatter is not due to measurement error.  In particular, single galaxies occupy tight regions in the $\dot{\Sigma}_* - \Sigma_g$ plane \citep[e.g.][]{bigiel2008,ostriker2010}. This suggests that there are additional parameters affecting star formation rates  \citep[e.g.][]{krumholz2012,saintonge2017}.   One avenue to explain galaxy-to-galaxy differences in star formation at fixed surface density is through the strong connections between pressure, dense gas fractions and star formation \citep[e.g.][]{wong2002,blitz2004,herrera2017,gallagher2018}. 
Galactic mid-plane pressure, in particular, can be linked to star formation rates \citep{ostriker2010, blitz2006}.

In normal spiral galaxies, old stars dominate the vertical gravity within disks.  Thus pressure-based analysis strongly links star formation rates to the total stellar mass and naturally explains why stars are likely to continue forming where stars have formed previously.  This motivates star formation laws of the form $\dot{\Sigma}_* \propto \Sigma_*^a\Sigma_g$ with $a \sim 0.5$ as in \cite{blitz2006}.  \cite{shi2018} find that a fit such as $\dot{\Sigma}_* \propto (\Sigma_*^a\Sigma_g)^b$, where $a=0.5$ and $b=1.09$ has significantly less scatter than a traditional Kennicutt-Schmidt relation for their observed data set.  

\cite{ostriker2010} attributed the bulk of star formation regulation to FUV, but did not independently treat turbulent support even though they recognized that it may be more important \cite[see also: ][]{herrera2017}.  Simulations have also shown that ISMs are not well characterized as just two simple, distinct phases \citep{kim2015,benincasa2016}.   These factors motivate revisiting the role of FUV with more comprehensive treatments.  

\subsection{Using simulations to study galactic star formation}
Isolated galaxy simulations are an ideal place to study the structure of the ISM and star formation; they offer both high resolution and the full galactic context without complicating factors.  There is a strong body of work studying the ISM using isolated galaxy simulations.  Most studies focus on the formation and evolution of GMCs, as the intermediate step between diffuse gas and star formation \citep[e.g.][]{tasker2009,benincasa2013,pan2015,rey2017,duarte2017,dobbs2018, pettitt2018}.  Other studies focus on the stability of gas, and searching for signatures of star formation \citep[e.g.][]{nguyen2018,agertz2015,grisdale2018,benincasa2016}.

Stellar feedback has historically been the main approach for regulating star formation in galaxy simulations.  Such feedback can be tuned to limit the amount of star-forming gas until star formation rates are acceptable.   A second, even simpler approach is to directly limit the efficiency of star formation, as was done in the AGORA isolated galaxy code comparisons \citep{agoraIC}.   When free parameter choices are exploited in this way, simply regulating star formation is not a meaningful result.  Simulations also have effective parameter freedom such as restrictions on the gas equation of state and resolution limits on gravity that effectively hold up the ISM.   These points argue for high resolution simulations with parameter-free, first-principles models as much as possible.  However, key processes such as star formation and early feedback phases are essentially impossible to resolve in galaxy simulations so some freedom in the models is unavoidable.

Current supernova feedback implementations often dramatically blow away gas from the simulated ISM.  This partly reflects the need to regulate the baryon content of entire galaxies so as to meet abundance matching constraints \citep{keller2016}.   While this may effectively regulate star formation in the galaxy as a whole, it hampers our ability to study the ISM-star formation connection in galaxy simulations.  

\subsection{The Role of FUV Heating}

FUV radiation is the dominant heating process for warm and cold neutral gas in the ISM.   It does this by ejecting electrons from dust grains which deposit their energy in the gas.  In the energy budget of a typical stellar cluster, FUV radiation provides nearly two orders of magnitude more energy than supernovae or stellar winds \citep[Starburst99,][]{leitherer1999}.  Ionizing UV heating is powerful but it is absorbed locally within star forming clouds which must be resolved before it plays a major role.   In contrast, because FUV is not absorbed by atomic hydrogen, it has a long mean free path, making it far-reaching compared to other types of feedback.

As noted above, \cite{ostriker2010} analytically explored FUV heating to regulate star formation.  The model did not include a distinct treatment of supernovae or feedback other than FUV.  Instead, it assumed turbulent pressure was proportional to thermal pressure which in turn was controlled by FUV.  Tying turbulence to FUV this way obscures the role for FUV as a regulator.   Remarkably, given its potential importance, there has been little other analytical work on star formation regulation taking FUV into account.

Different forms of support, such as warm gas due to FUV, hot gas due to supernovae and turbulence behave differently within the ISM.  Whereas gas that has been heated by supernovae can leave the disk in an outflow, FUV heated gas remains in the ISM. The role FUV plays in regulating star formation cannot be disentangled from its role in maintaining the structure of the ISM.  The necessity of producing a realistic ISM is a powerful constraint that should help us understand the roles of different feedback.   

Combining both supernovae feedback and FUV heating provides simulations with a new dimension to explore.  However, FUV heating is not commonly employed as a feedback.  If radiative transfer is not available, FUV radiation can be included as a prescribed background heating rate \citep[e.g.][]{kim2011,benincasa2016,hu2017} but this is not self-consistent. It does not focus heating where stars are being formed both spatially and temporally.  For very small volumes, a uniform FUV field can be tied to the star formation rate as in \cite{kim2011}.  

Numerical works have begun to include radiation fields derived from the stellar sources for entire galaxy simulations.  \citet{forbes2016} studied the effect of photoelectric heating on dwarf galaxies under the assumption that they were entirely optically thin to FUV due to their low metallicity.  They conclude that FUV is a very effective stellar feedback in these small galaxies ($\lesssim 0.01 $ the Milky Way in total mass). 

Typical galaxies like the Milky Way are optically thick to all UV radiation.  
The mean free path for FUV in the ISM is $\sim 1$ kpc which means sources up to $\sim$ 10 kpc away may contribute to the local flux (see appendix). One approach is to model this with local absorption estimators. \citet{hopkins2012} and related papers \cite[e.g.][]{onorbe2015} employ a Sobolev-like approximations, where gas column is estimated as $\Sigma \sim \rho L$ with a characteristic length $L = \rho/|\nabla \rho|$.   This approximation was designed for radiation escape or self-shielding estimates at the receiving gas \cite[e.g.][]{gnedin2009}.  
Instead, these simulations assume an optical depth associated with the easiest escape path at the source, assumed to be $-\nabla \rho$.  Starlight experiences a fixed attenuation regardless of whether it travels in this direction or not.  Similarly, setting $L$ equal to the Jeans length \citep{diemer2018,  marinacci2017,  lagos2015} assumes the local gas is smooth, bound and the only source of absorption. It also prevents directional variations.

Even simple galactic radiative transfer models require directional dependence \cite[e.g.][]{byun1994}.  \citet{wang2018} show that UV extinction varies by a factor of $\sim 4$ depending on inclination for galaxies at intermediate redshifts.  Local absorption approximations are unlikely to be accurate except for photons leaving vertically (e.g. face-on mock images).   To study photoelectrc heating, we must follow FUV radiation moving laterally within the galaxy, where gas typically sees photons that have traversed several kpc of ISM structure.  Self-consistent FUV heating via full radiative transfer is the best way to simultaneously quantify its roles in regulating star formation and setting the structure of the ISM. 

A promising approach to full radiative transfer is the M1-closure \citep{aubert2008} which uses an explicit flux variable to improve the directionality and diffusiveness of prior moment methods such as flux limited diffusion.  M1 still merges crossing rays and cannot quite match detailed ray tracing such as VET \citep{davis2014},
but should be reasonable for average intensities required for heating.  A common associated choice is a dramatically reduced speed of light to allow larger time steps.   
M1 has been added to several major codes \citep{rosdahl2013, kannan2019, hopkins2020}.  \cite{rosdahl2015} ran a high-resolution galaxy with multi-band radiative transfer but did not include a FUV band.   
\cite{hopkins2020} have recently run M1-radiative transfer on galaxies including an FUV band.  Their basic results do not show FUV to be an important regulator.  However, they have not yet presented detailed results on FUV.  Their results also assume relatively high opacities, so may be taken as a lower limit on the possible effect of FUV.

In the current work, we use the TREVR \citep{grond2019} radiative transfer algorithm to implement self-consistent FUV heating in simulations of entire galactic discs. This is a ray-tracing approach and thus distinct from moment-based approaches like M1 and thus complementary to existing work.

In Paper I \citep{benincasa2016}, we explored the coupling between the ISM, its pressure, star formation and feedback.  The high-resolution, isolated galaxy models employed there used specified FUV fields, a parameter-driven supernova model and a fixed galactic potential without an old stellar disk. Paper I demonstrated that time-averaged pressure balance is a key feature of well-resolved, simulated galaxies.  It also established strong connections between properties such as scale height, gas surface density and star formation rates.  In that work supernovae were the primary source of pressure support and FUV was not self-consistently treated. The isolated galaxy of Paper I also did not have an old stellar disk which prevented direct comparisons to similar nearby galaxies. The current work corrects these oversights.

\textit{In this paper, we explore the impact of self-consistent FUV heating from radiative transfer in combination with physically well-constrained supernova feedback on the ISM and star formation in an isolated galaxy with a live halo and old stellar disk.}
The remainder of this paper is laid out as follows:  In section \ref{sec:model}, we describe our chosen galaxy model, a modification of the isolated disk test cases used in the \textsc{Agora} comparison project. In section \ref{sec:methods} we describe the model for radiative transfer that we employ in this work. We then contrast different combinations of feedback choices and their roles in setting different galaxy and ISM properties in sections \ref{sec:props0} and \ref{sec:phases}. 

\section{Galaxy Model} \label{sec:model}

Our isolated galaxy model is a higher-resolution version of the initial conditions from the \textsc{Agora} High-resolution Galaxy Simulations Comparison Project \citep{agora,agoraIC}.  The galaxy has a live stellar disk and bulge, as well as a live dark matter halo. The dark matter halo has an NFW density profile \citep{navarro1997}.  The halo has $M_{200}=1.074\times10^{12}$ M$_{\odot}$ and a halo concentration parameter, $c=10$.  The stellar disk has an exponential density profile with a total mass of $3.438\times10^{10}$ M$_{\odot}$.  The stellar bulge has a total mass of $4.927\times10^9$ M$_{\odot}$ and follows a Hernquist profile \citep{hernquist1990}.

We began with the standard (low)  \textsc{Agora} resolution initial condition, but then split each particle 64 times to increase the resolution. The dark matter halo has 6.4 million particles, each of mass $1.956\times10^5$ M$_{\odot}$. The stellar disk and bulge combined have 7.2 million particles, each of mass 5360 M$_{\odot}$. The gas disk is composed of 6.4 million particles, each of mass 1342 M$_{\odot}$. We employ the standard Gasoline piecewise polynomial gravitational softening, with a softening length of 80 pc (so that force is Newtonian at a distance of 160 pc).

\begin{figure}
	\begin{center}
		\includegraphics[width=0.45\textwidth]{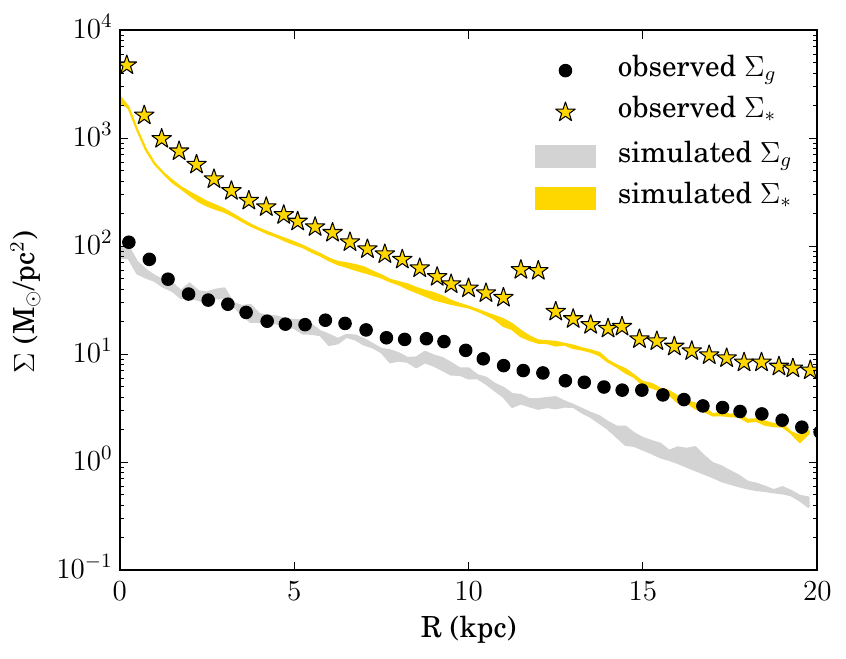}
	\end{center}
	\caption{A comparison of our simulated galaxy surface density to the gas surface density of the sunflower galaxy, NGC 5055.  The grey bar shows the range of simulated $\Sigma_g$ and the yellow bar the range of simulated $\Sigma_*$, measured after 200 Myr of evolution. The filled circles show the total gas surface density, $\Sigma_{HI+H_2}$ as reported in \citet{bigiel2008}. The filled stars show the total stellar surface density as reported in \citet{leroy2008}. Our galaxies show good agreement with NGC 5055, within 30\%, until the outer regions of the disk (R $\gtrsim10$ kpc).}
	\label{fig:sd}
\end{figure}

We picked the public \textsc{Agora} galaxy IC to facilitate comparisons with other work, including the original study \citep{agoraIC} and newer work that also uses it \citep[e.g.][]{agertz2015,grisdale2017,grisdale2018,semenov2017,semenov2018}.  However, we note that this galaxy has a significantly higher surface density than typically estimated for the Milky Way, its ostensible target \citep[e.g.][]{nakanishi2016}.   It is still in the typical range for normal spiral galaxies.  In fact, the \textsc{Agora} IC has similar characteristics with NGC 5055, the sunflower galaxy. 

In Figure \ref{fig:sd} we plot a comparison of the gas and stellar surface density in our simulations to those of NGC 5055. The total gas surface density is a combination of THINGS HI \citep{walter2008} and H$_2$ as measured by HERACLES CO \citep{leroy2009}. The stellar surface density is from the GALEX nearby galaxy survey as reported in \citet{leroy2008}. Both of the model profiles are similar to the observational data within 10 kpc.  The observed stellar surface density is $\sim30\%$ higher than in our galaxy. These radial distributions, especially the stellar distribution, are potentially important quantities in determining the outcomes of galaxy scaling relations \citep{ostriker2010}.

For comparison, paper I \citep{benincasa2016}, used a static potential with no old stellar population.  The presence of the heavy, old stellar disk has the effect of stabilizing the gaseous disk, but also acts as a driver for spiral structure.  Locally, the old stellar disk is an important source of vertical gravity to limit the gas scale height.

\section{Methods} \label{sec:methods}

\begin{figure*}
	\begin{center}
		\includegraphics[width=0.95\textwidth]{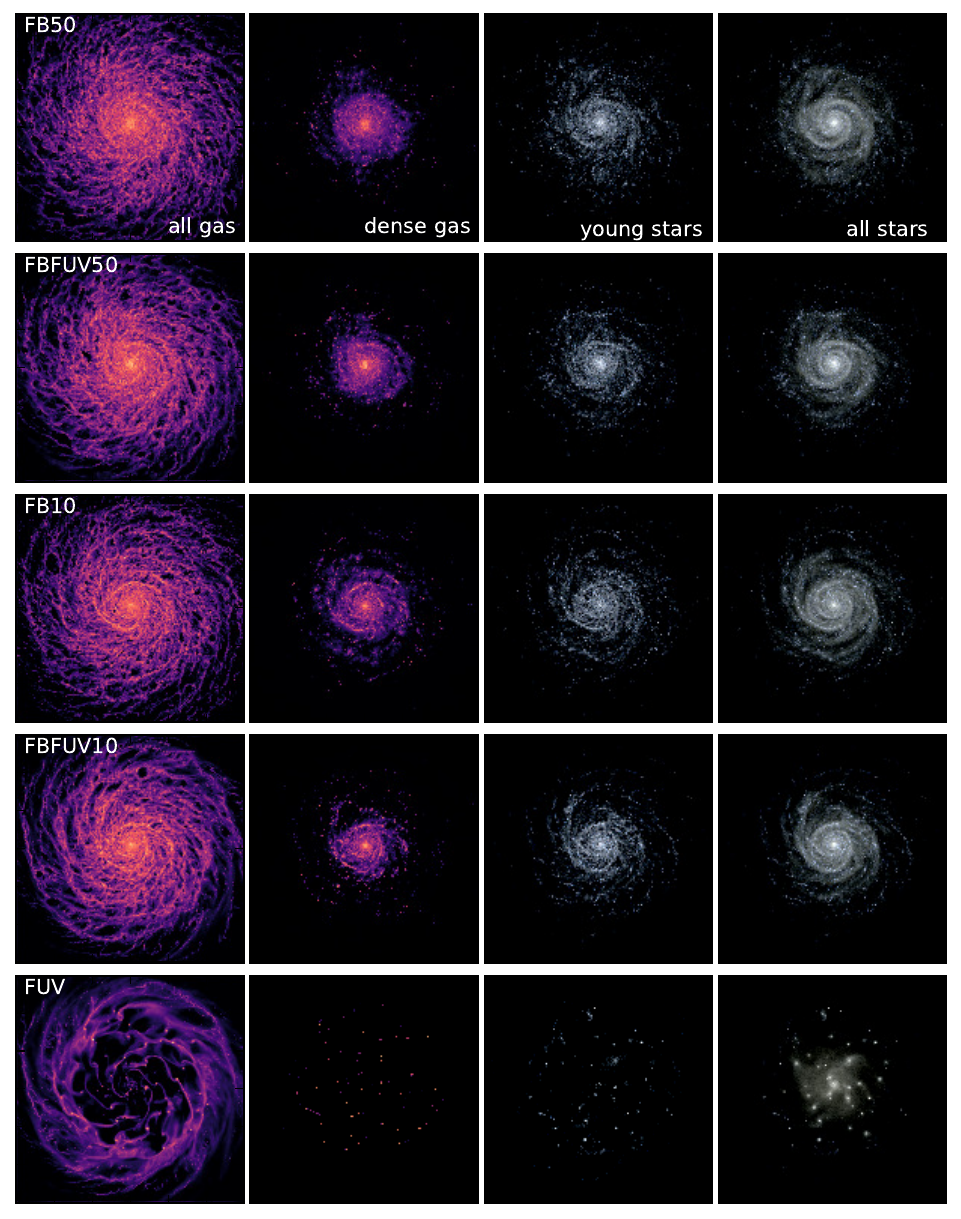}
	\end{center}
	\caption{Gas and stellar maps for the galaxies in the simulation suite. \textit{Left:} Total gas surface density. \textit{Middle-Left:} Dense gas surface density, which here means any gas with $n>100$ cm$^{-3}$}. In the gas surface density columns, the colour-scale runs from 1 M$_{\odot}$/pc$^2$ (black) to 500 M$_{\odot}$/pc$^2$. \textit{Middle-Right:} Young stellar luminosity; here young stars have formed less than 100 Myr in the past. \textit{Right:} Stellar luminosity for all stars formed since the start of the simulation. When both super-bubble feedback and radiative transfer are included there is very little difference in these maps from galaxy to galaxy. However, it is clear tof orderhat FUV heating alone cannot regulate star formation and sustain galactic disk, as seen in the final row.
	\label{fig:faceproj}
\end{figure*}

We simulated a suite of different feedback treatments, as listed in Table \ref{tab:simtable}, on our galaxy model to study star formation and ISM evolution over a period of 400 Myr.  This allows time for the galaxy to settle into a well-regulated state if one exists.  These feedback treatments and other aspects of the simulations are described in this section.

We use the modern SPH code \textsc{Gasoline} \citep{wadsley2004,wadsley2017}. As demonstrated in \cite{wadsley2017}, modern SPH, as implemented in \textsc{Gasoline}, performs particularly well for supersonic, turbulent gas such as that present in disk galaxies. The simulations employ star formation, feedback and metal cooling following the standard prescriptions presented in MUGS2 \citep{keller2015}.  These models have been successful in producing realistic disk galaxies, including stellar content over cosmic time, in cosmological zoom-in simulations \citep{keller2016}.  A new component in this work is the inclusion of radiative transfer to implement self-consistent FUV radiation from young stars throughout the entire disk.

\subsection{Radiative Transfer}

Outside hot superbubbles, the interstellar medium is primarily heated by the photo-electric effect on dust grains due to FUV radiation \citep{wolfire2003}.  X-rays and cosmic rays are of secondary importance except at very low densities or in dense, shielded gas.  Gas-ionizing UV radiation is of comparable importance in principle, except that the high optical depths limit it to HII regions occupying a fraction of a percent of the ISM.  Thus when studying the typical ISM ($0.01-100$ cm$^{-3}$, $100-10000$ K),  FUV is the most important heating process and it sets the temperature.

In this study, we fully propagate the FUV flux emitted by young star clusters estimated using integrated spectra from Starburst99 over the band 912-1550\,\AA{} \citep{leitherer1999}.  This is the first time this has been done on-the-fly using ray-tracing with absorption in a full, isolated galaxy disk simulation. The radiation field is modelled as a single FUV band as discussed in \cite{grond2019}.  This band heats the gas via the photo-electric effect on dust which is assumed to be linear in the local gas metallicity with a $3 \%$ heating efficiency (a fairly typical value, \cite{tielensBook}).  We employ an opacity due to dust absorption of $300$ cm$^2$/g (scaled linearly from solar metallicity).  We note that the Lyman-Werner band, which dissociates molecular hydrogen, is also in the FUV range.  However, the current work does not treat molecular gas formation or destruction.

The simulations presented here employed the original version \citep{woodsPHD} of the radiative transfer algorithm, TREVR, detailed in \cite{grond2019}.  TREVR (Tree-based REVerse Raytracing) is a novel algorithm for computing radiation fields.  TREVR estimates local, angle-averaged intensities for heating and chemistry based on reverse ray-tracing back to available sources (which are merged using the tree) and includes absorption by intervening gas and dust (but not scattering in current versions).  The original TREVR algorithm did not use adaptive accuracy controls.  It slowed the code by roughly a factor of two when self-consistent FUV was being computed.  The final version of TREVR (\cite{grond2019}, not used in the calculations presented here), includes detailed error control via refinement at the expense of considerable extra computation.  As such, it would not be feasible (at present) to use for full simulations but is useful for generating highly accurate radiative transfer results for comparison at a specific output time.  In the Appendix, we use it to show that the fast, original scheme is accurate for FUV, particularly when it comes to ISM temperatures.

This work recreates a typical Milky Way-like ISM similar to that of \cite{wolfire2003}, who modeled heating rates and ISM phases as a function of Galactic radius.  For fixed opacity, clumping in radiative transfer tends to reduce the absorption relative to a uniform medium, as most rays miss denser structures.   Wolfire's FUV model employed a relatively low opacity value ($\sim$ 300 cm$^2$/g without scattering) to be used with smoothed, average ISM densities (e.g. without molecular clouds or smaller-scale structures).  The current work has intermediate resolution, with limited ability to model denser gas $\gtrsim$ 100 cm$^{-3}$ and thus clouds-scale clumping effects on radiative transfer are not present.  Given these considerations, we argue that 300 cm$^2$/g is a low but reasonable opacity.  Thus this work could be considered an upper bound on the effect of FUV. 

For other radiation fields and heating processes (including EUV and cosmic rays), we used fixed background rates as in paper I \citep{benincasa2016}. Together with the density and temperature, these determine the local, non-equilibrium ionization state of the gas.  We note that photo-ionization and photo-dissociation due to local radiation sources directly impacts a tiny portion of the ISM, on parsec scales.  At the current resolution, we do not capture such dense star forming sub-clumps ($\gtrsim 10^4$ cm$^{-3}$).  Our star-forming clouds (100-1000 cm$^{-3}$) are readily destroyed by the stellar feedback processes described below.

\begin{table}
	\caption{List of simulation details}
  	\label{tab:simtable}
	\begin{center}
  	\begin{tabular}{lccccc}
    		\hline
		name & SNe & FUV & cosmic rays & $E_{SN}$ (ergs) \\
		\hline
		FB50 & \cmark & \xmark & \cmark & $5\times10^{50}$ \\
		FUVFB50 & \cmark & \cmark & \cmark & $5\times10^{50}$ \\
		FB10 & \cmark & \xmark & \cmark & $10^{50}$ \\
		FUVFB10 & \cmark & \cmark & \cmark & $10^{50}$ \\
		FUV$^{\star}$ & \xmark & \cmark & \cmark & none  \\
  	\end{tabular}
	\end{center}
\end{table}

\begin{figure*}
	\includegraphics[width=0.93\textwidth]{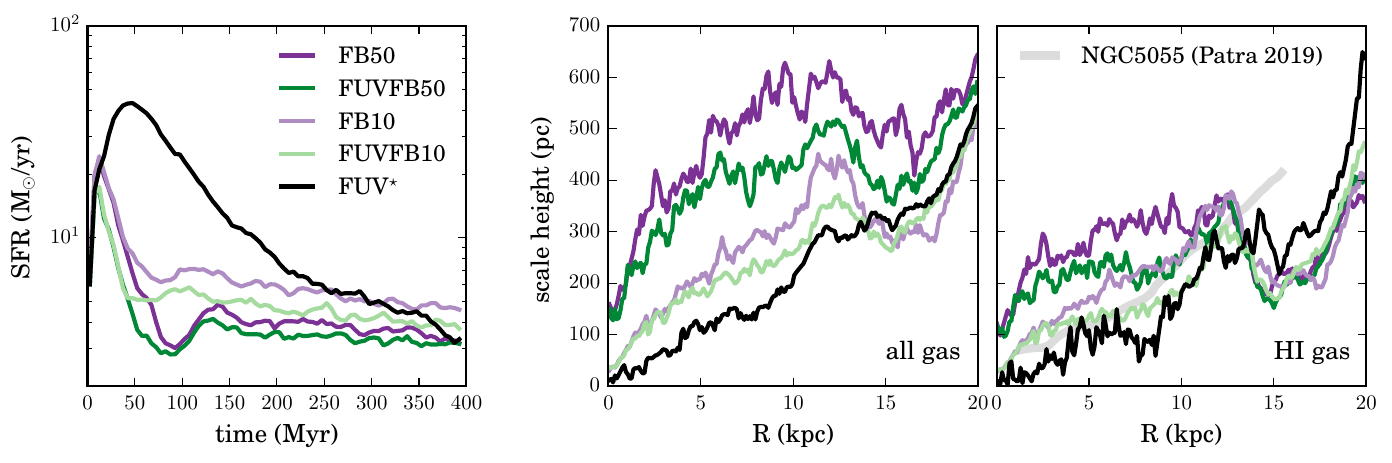}
	\caption{A comparison of the global SFR and the gas scale heights for the galaxies in the suite. \textit{Left}: The global star formation rate. In all of the galaxies that include both super-bubble feedback and FUV heating, there are only small changes in the star formation rate.  However, when the only form of feedback is FUV heating, there is a dramatic increase in the amount of star formation. This galaxy can only decrease its star formation rate by consuming a significant amount of the star-forming fuel. \textit{Middle}: The scale height for all gas in the galaxies. In this case the difference between including only FUV heating and both modes of feedback is less stark in contrast. As expected, cases with stronger super-bubble energy have larger scale heights. Interestingly, including the radiative transfer on top of super-bubble leads to a decrease in the scale height. \textit{Right}: The scale height for HI gas, the distribution as quoted in \citet{patra2019} is plotted in grey for comparison. The case with FUV heating and a lowered feedback efficiency, FUVFB10, shows the best agreement while still regulating star formation. The scaleheight shown here is measured at 400 Myr.}
	\label{fig:sfrH}
\end{figure*}
\subsection{Feedback}
We employ the super-bubble feedback method as detailed in \cite{keller2014}.  This model includes the effect of electron conduction to model the combined mechanical output of young clustered stars.  A self-similar solution was first presented by \cite{weaver1977}.   Super-bubbles form when fast outflows from massive stars (such as supernovae) shock and merge into an expanding hot bubble.  The bubble pressure sweeps up a cold shell of ISM.  Electron conduction regulates the interior mass and temperature ($\sim 3\times 10^6$ K) of the bubble.  The steep power-law dependence of conduction on temperature makes the model insensitive to the conduction coefficient (e.g. effects due to tangled magnetic fields) so there are effectively no free parameters.  The bubble's behaviour is strongly determined by how much energy is injected and the density of the medium, which are easily determined in simulations.  The model includes conduction, evaporation of cool into hot gas and a sub-grid phase representing the swept up cold gas shell.  The sub-grid phase is short-lived as evaporation rapidly generates well-resolved hot bubbles.  Both phases cool normally.  These properties make the outcomes insensitive to resolution \citep{keller2014}.

Commonly-employed prior methods, such as that used in \cite{benincasa2016}, have several free parameters (e.g. a parameter for an energy transfer timescale or a cooling time and a parameter to determine how much mass is affected).  Together with resolution these affect basic outcomes such as feedback temperatures.  This reflects the fact that the structure of the ISM has typically been a secondary concern as long as the overall star formation rate was acceptable.  Prior work showed that the strong supernovae feedback was highly destructive to the ISM in a manner that suggested too much of the kinetic energy was being deposited there.  One long-standing approach is to hydrodynamically decouple feedback material from the ISM in galaxy simulations \citep[e.g.][]{springel2003}.  

We would like to tightly constrain the role of each feedback.   However, for supernovae in particular, there is uncertainty regarding how much energy directly affects the ISM \citep{gentry2020}.  Recently, \cite{elbadry2019} explored super-bubbles with high-resolution, one-dimensional simulations and found that additional energy could be lost during evaporation.  In addition, \cite{keller2016}, found that $50\%$ of the supernovae energy going into the super-bubbles was sufficient to correctly simulate the baryon content of Milky Way-like galaxies and their progenitors over cosmic time.  We take this as an energy upper limit.  However, as noted above, strong coupling to the ISM by supernovae outflows is an unavoidable consequence of limited resolution.  An equivalent statement of this issue is that the effective Reynold's number of simulations is generally much lower than that of the real ISM so that boundary layers and viscous coupling are over-estimated.  A crude way to model this is to assume some part of the feedback energy vents directly to the galactic halo, as in \cite{springel2003}, and that such losses do not immediately affect the ISM. 

For these reasons, we use two different supernovae feedback strengths, 50\% and 10\% of the maximum value, or $5\times10^{50}$ ergs and $10^{50}$ ergs respectively, to attempt to bracket the range of possibilities.

\subsection{Star Formation}
We use a common star formation prescription, where stars form following a Schmidt law:
\begin{equation}
\frac{\text{d}\rho_*}{\text{d}t} = c_* \frac{\rho_g}{t_{dyn}},
\end{equation}
where $\rho_*$ is the density of new stars formed, $\rho_g$ is the density of eligible gas, $t_{dyn}=1/\sqrt{4\pi G \rho_g}$ is the dynamical time and $c_*$ is the chosen efficiency. Gas is considered eligible for star formation if it lies above a set density threshold, lies below a maximum temperature and belongs to a converging flow. This is a typical star formation method \citep[e.g.][]{katz1992}.  The small fraction of particles that are currently in a two phase state cannot form stars.  We assume that these particles, near recent star formation events, are analogous to unbound GMCs. In this study we employ a $c_*$ of 0.05, a density threshold of 100 cm$^{-3}$ and a maximum temperature threshold of 1000 K. 

\section{Results}\label{sec:props0}
\subsection{Global Galaxy Properties} \label{sec:props}

We begin by considering the basic properties of the galaxies in our suite after 400 Myr of evolution (detailed in Table \ref{tab:simtable}). As a first diagnostic, we consider the visual appearance of the galaxies. Figure \ref{fig:faceproj} shows face-on images for different quantities of interest. The left-most panel shows the total gas surface density of the disk.  The second panel shows the gas surface density of the disk when considering only gas above 100 cm$^{-3}$, the analogue of molecular gas in our simulations.  The third panel shows a synthetic stellar map for young stars.  To show the primary sources of FUV, we show only stars that have formed in the last 100 Myr \citep{salim2007,murphy2011,kennicutt2012}. Finally, we show a synthetic stellar map for all stars, to show how the stellar disk has built up over time in each case: this includes only stars that have formed since the start of the simulation and excludes the initial stellar disk.

If we consider each of the gas panels in Figure \ref{fig:faceproj}, the cases that have both super-bubble and FUV feedback are qualitatively similar: in each case we see a flocculent spiral with gas that extends far beyond the extent of the star-forming region of the galaxy. The cases with a lower SNe feedback energy, FUVFB10 and FB10, have more apparent spiral structure than those with higher feedback energy. This manifests itself in the stellar disk as narrower spiral arms. There is, however, a stark contrast when we consider the case with only FUV heating and no other form of feedback. In the gas we can see that there is a high degree of fragmentation for this galaxy. This results in highly clustered star formation and a dense stellar nugget at the galaxy's core, as well as over-consumption of gas in the galactic disk

The global star formation rate (SFR)  for each of the galaxies is plotted on the left side of Figure \ref{fig:sfrH}. The SFR provides us with concrete evidence of what is suggested in the maps in Figure \ref{fig:faceproj}. Firstly, FUV heating alone cannot regulate star formation in the galaxy. When FUV heating is the only source of feedback, the star formation rate is very high and only begins to decrease as the galaxy runs out of gas (black line).  This is similar to no feedback cases in this and the mode of excessive star formation via large gas clumps.  

Simulations with supernova feedback stabilize their SFR by $\sim200$ Myr.  
The cases with super-bubble feedback only are plotted as purple lines, while cases with FUV heating and super-bubble are plotted in green. As expected, decreasing the feedback energy results in more star formation \citep[see paper I:][]{benincasa2016}. Further, adding FUV heating on top of super-bubble feedback results in a small decrease in the SFR, of order 30\%.

Another notable feature in these star formation rates is how quickly the star formation rate settles or dips after the initial burst. The cases with both modes of feedback (green lines) are able to regulate and end the burst more quickly than the cases with only super-bubble feedback (purple lines). This timing difference is approximately 10 Myr.  In our chosen feedback model, supernovae do not begin until 4 Myr after a star is born. FUV radiation however, begins heating the surrounding environment immediately after the star is born. This means that regulation can begin more quickly.

Next, we compare the gas scale heights for each of the galaxies, shown on the right side of Figure \ref{fig:sfrH}. In this figure we plot both the total gas scale height (left) and the HI gas scale height (right). For comparison, we also plot the HI scale height for NGC 5055 as quoted in \citet{patra2019}.  We find that the galaxies with the most supernova feedback energy have the highest scale heights.  The response is some what sub-linear, as shown in Paper I due to the dependence of gravity on height in disks.

The SN energy injection rate is proportional to the star formation rate multiplied by 5 for the FB50/FUVFB50 cases, 1 (unchanged) for the FB10/FUVFB10 cases and zero for the FUV-only case.  The inclusion of FUV heating leads to a decrease in the scale height because the warm medium it generates provides pressure support (discussed more in Section~\ref{sec:phases}).  This allows for a decreased role for super-bubble feedback in supporting the gas and the lower star formation rate.  Super-bubble feedback energy is responsible for inflating the disk and causing outflows from the ISM. 

\begin{figure*}
	\includegraphics[width=0.9\textwidth]{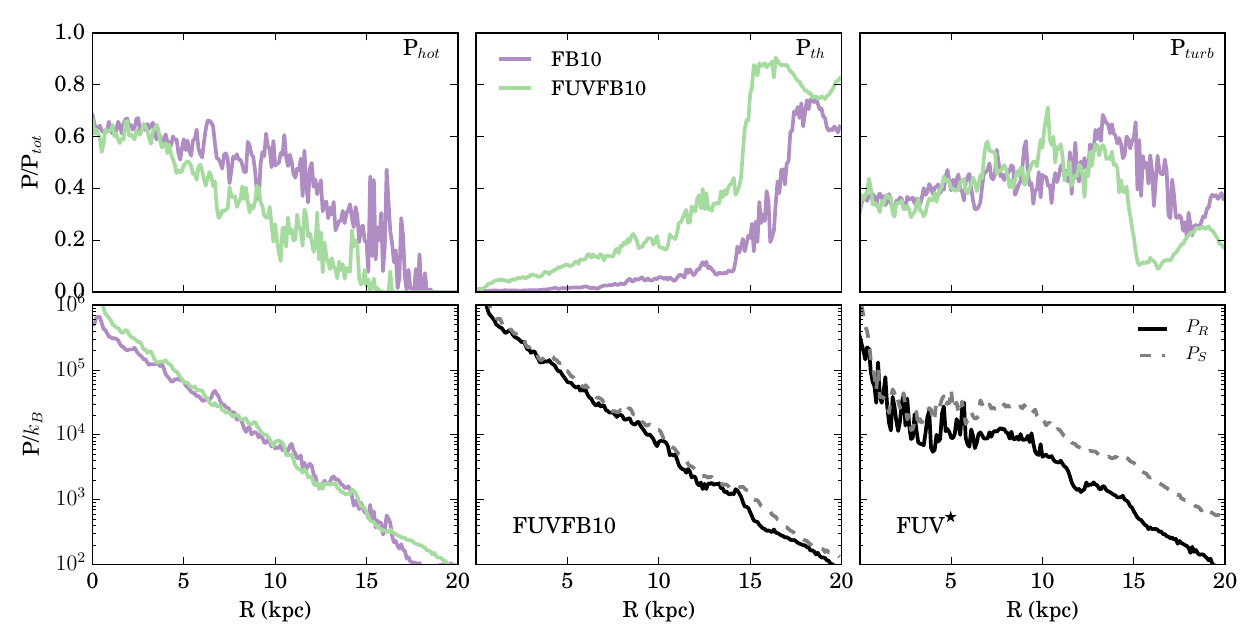}
	\caption{Comparison of the division of pressure in different means of support.  \textit{Left}: Contribution in each galaxy to P$_{hot}$. \textit{Middle}: Contribution in each galaxy to P$_{th}$. \textit{Right}: Contribution in each galaxy to P$_{turb}$. Adding FUV heating does not change the amount of pressure held in the turbulent component. However, it does lead to more support coming from a thermal component in the ISM. We have plotted only our lower feedback galaxies for clarity, but these trends hold for the FB50 cases as well.}
	\label{fig:pressure}
\end{figure*}

\subsection{Pressure Balance in the ISM}{\label{ssec:pressure}}

Pressure facilitates the regulation of star formation in galaxies and it is connected to the dense gas fraction \citep{blitz2004, blitz2006, ostriker2010}.  We explore the pressure balance in our simulated galaxy disks in this section, following the approach of Paper I.  

The star formation rate in a galaxy is set by the balance between the pressure required and the pressure support. The level of pressure required is set by gravity in the disk:
\begin{align}
 \mathrm{P}_{\mathrm{R}} &= \mathrm{P}_{dm} + \mathrm{P}_g + \mathrm{P}_* \nonumber \\ 
&= \frac{1}{2}\Omega^2\Sigma_g \mathrm{H}_g  +  \frac{1}{2}\pi G \Sigma_{\text{g}}^2 + \pi G\Sigma_g\Sigma_* \left(\frac{\mathrm{H}_g}{\mathrm{H}_g + \mathrm{H}_*} \right)  \label{eqn:ptotcomp}
\end{align}
where $\Omega$ is the rotation rate, $\Sigma_g$ is the gas surface density, $\mathrm{H}_g$ is the gas scale height, $\Sigma_*$ is the stellar surface density and $\mathrm{H}_*$ is the stellar scale height.  The dark matter halo primarily sets the rotation rate and the vertical component is proportional to $\Omega^2$.  See \citet{benincasa2016} for a detailed derivations of these terms. This required pressure, $P_R$, is the pressure needed to support the weight of the ISM in a state of hydrostatic equilibrium.

The pressure support we measure in the disk is calculated by:
\begin{align}
  \mathrm{P}_{\mathrm{S}} &= \mathrm{P}_{th} + \mathrm{P}_{hot} + \mathrm{P}_{turb} \nonumber \\
    &= \frac{\Sigma_{\mathrm{g}}}{2\,H_g} \left(
\frac{2}{3}u_{\mathrm{th}} + \frac{2}{3}u_{\mathrm{fb}} + v_z^2
\right)_{z=0}. \label{eqn:ptotsupport}
\end{align}
This is the mid-plane support and so all these quantities take on their
mid-plane values.  The mid-plane density, $\rho_{\mathrm{g},0}$, is well
approximated by the gas surface density divided by twice the gas scale height,
$H_g$.  $\mathrm{P}_{th}$ refers to warm gas primarily heated by FUV radiation whereas $\mathrm{P}_{hot}$ refers to supernovae heated gas in the mid-pane.  In Paper I only the hot component was directly linked to star formation. In the current simulations, FUV is linked not only to local star formation but young stars within several kpc via radiative transfer.

In Figure \ref{fig:pressure} we plot a selection of pressure quantities of interest. To discern the relative importance of each, we plot these as ratios with the total pressure in the top row We note that while here we plot a comparison of the FB10 cases, the trends discuss hold also for the higher feedback, FB50, cases as well. The top-left of Figure \ref{fig:pressure} shows the amount of support from $P_{hot}$, the top-middle shows the amount from $P_{th}$ and the top-right shows the amount from P$_{turb}$. For clarity, on the two of the galaxies in the suite are plotted, FUVFB10 (green) and FB10 (purple). As expected, adding the FUV heating decreases the role for super-bubble feedback in supporting equilibrium (P$_{hot}$). There is very little impact on the turbulent pressure support in the disk.

In the bottom row of Figure \ref{fig:pressure} we plot the total pressure, as well as the state of pressure equilibrium in two sample disks. In the bottom-left panel the total pressure support for FUVFB10 and FB10 are plotted. The total pressure is equal, so we can see that changing the types of feedback merely changes the partitioning of the pressure into different forms of support.

The role of FUV radiation is to heat cold neutral gas. When FUV radiation is included, as in case FUVFB10, a larger fraction of the pressure support is held as a thermal component (top-middle panel of Figure \ref{fig:pressure}). In order for the total galactic pressure to remain the same, this must be made up by super-bubble energy, or $P_{hot}$. This manifests itself as changes in the phase structure of the ISM, as will be shown in section \ref{sec:phases}.  This demonstrates that FUV is critical to support the ISM in the outer parts of galaxies.

The final two plots show the state of pressure equilibrium in FUVFB10 and FUV$^{\star}$. What is plotted here is different from what is plotted in \citet{benincasa2016} in the sense that we do not have to rely on time-averaging to see the differences. In these simulations we have an old stellar disk whose gravity dominates the required pressure component.  In paper I, gas determined the dynamics of the galaxy, which made it hard to link the simulations to nearby galaxies.

This picture is different when we consider the case with FUV heating and no super-bubble feedback, FUV$^{\star}$.  This is the only case where we do not see global pressure equilibrium. The imbalance is already in place after 50 Myr.  FUV heating by itself is unable to regulate the star formation rate. This results in the fragmentation of the gas into dense knots and a very high level of gas consumption. This significantly alters the distribution of gas throughout the disk.  FUV heating due to star formation is more intense in the knots but is not effective as a feedback because it cannot unbind them.   However, the scale-height for this galaxy, shown in Figure \ref{fig:sfrH} is smaller than all of the other galaxies: less than 100 pc in the inner regions of the disk. At this scale-height, we are approaching the resolution limit set by the gravitational softening.  Thus, this pressure disagreement requires further investigation at higher resolution.  We note that in \textsc{gasoline}, hydrodynamic resolution is not limited so it favours support over collapse below the gravity resolution scale.

\begin{figure*}
	\includegraphics[width=0.98\textwidth]{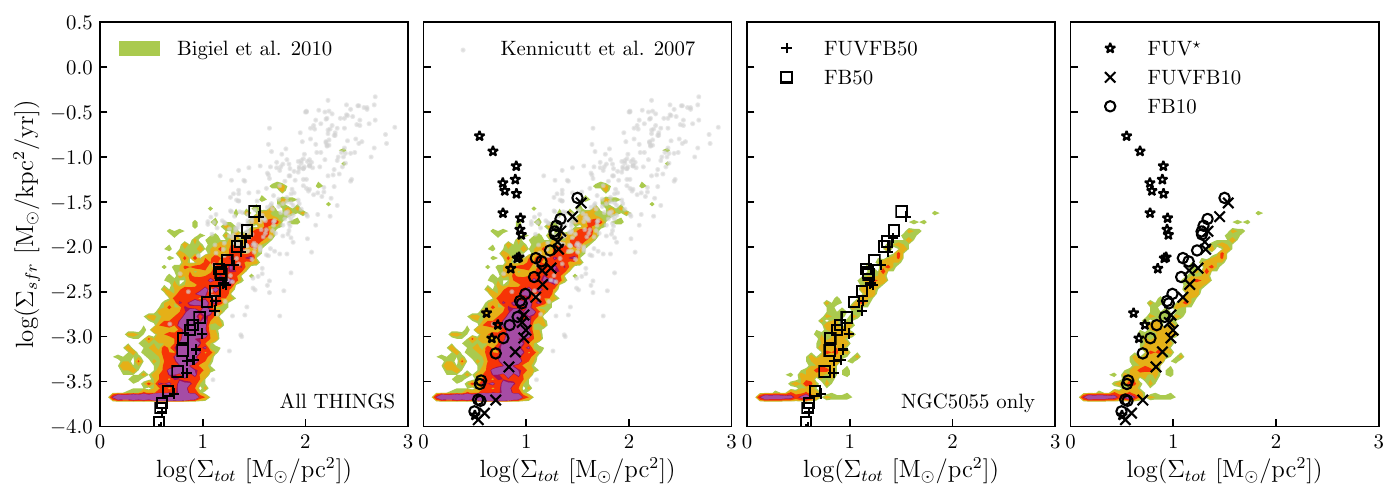}
	\caption{The Kennicutt-Schmidt relation for our simulated galaxies compared to observational samples.  The black symbols show the five galaxies in our suite.  For our data we plot only values for radii outside 2.5 kpc.  In the left-most two plots the coloured contours show the space occupied by local measurements for all the THINGS sample; in the right-most two plots the contours show the space occupied by NGC 5055 \citep{bigiel2008}.  The dots in the top row show the local measurements in M51 from \citet{kennicutt2007}.}
	\label{fig:kshi}
\end{figure*}

\begin{figure}
	\includegraphics[width=0.45\textwidth]{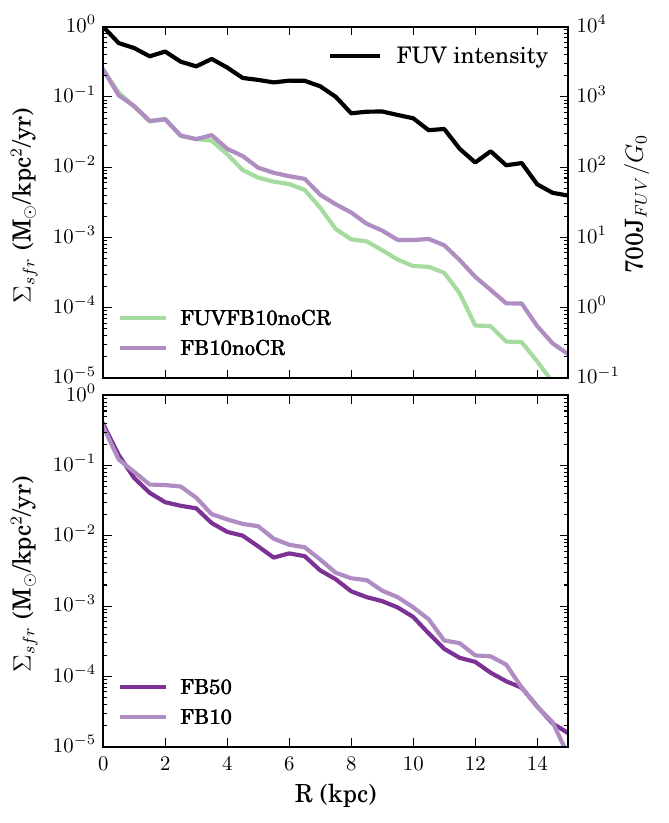}
	\caption{The SFR surface density as a function of radius for four galaxies in the suite. The inclusion of FUV radiation changes the slope of the SFR surface density relation, as shown in the top plot. In contrast, increasing the amount of energy from super-bubble feedback changes the normalization of the SFR surface density curve but not the overall slope, as shown in the bottom plot. We posit that FUV radiation has the highest impact in the outer disk of galaxies. FUV radiation has a long mean free path and thus can impact the ISM in a more long range way. This is shown by the FUV intensity, plotted as the black line in the top panel. This line does not share the same slope as the SFR surface density: if the feedback were localized they would share the same slope.}
	\label{fig:truncation}
\end{figure}

\subsection{The distribution of star formation}

The Kennicutt-Schmidt relation is an empirical relation used to characterize the star-forming ISM in galaxies. {\bf It is important to note that the spread in this relation is real, as explored by \citet{leroy2008} and others.  It cannot be accurately characterized with a single power-law and depends on individual galaxy properties beyond the local gas surface density.  Thus when we say KS relation, we are referring to the specific relationship between star formation rates and gas surface densities for an individual galaxy and not a specific power-law fit.}  In Figure \ref{fig:kshi} we plot the Kennicutt-Schmidt relation for the simulated galaxies and compare to a selection of observational data. The contours in Figure \ref{fig:kshi} show locally measured data from two different surveys. The coloured contour shows the space occupied by the THINGS galaxies as reported in \citet{bigiel2008}; these measurements are taken at 750 pc scales. The small grey circles show local measurements of M51 as reported in \citet{kennicutt2007}; these measurements are taken on 300 pc scales.  In the left two plots we compare to the entire sample of galaxies, whereas in the right two plots we compare specifically to NGC 5055. The simulated galaxies are plotted as the different open symbols. These measurements were taken in 500 pc radial annuli, to agree with the resolution of the radial measurements reported in \citet{leroy2008}.  For the star formation rate surface density we consider only the stars that have formed in the last 100 Myr. Additionally we take only measurements outside a galactic radius of 2.5 kpc.

Generally, our simulated data agree with the observational trend for cases with supernova feedback.  The relation for the FUV$\star$ case further shows that FUV heating along cannot regulate the star formation rate.  There the relation between $\Sigma_g$ and $\Sigma_{sfr}$ is nearly vertical, such that pieces of the galaxy at the same $\Sigma_g$ can have star formation rates that vary by several orders of magnitude. This is in contrast to \cite{ostriker2010} who suggested that FUV can be the main regulator of star formation in the ISM.
 
For cases that have both types of feedback, the fit to NGC 5055 is quite close.  In particular, we are encouraged by the fact that the KS relation is much tighter for this individual galaxy than the large spread in the entire population as is to be expected for any individual galaxy. 

On closer inspection there is an offset at higher surface densities such that the slope is somewhat steeper.  Given the large variation in slopes for individual galaxies with the overall distribution \citep[e.g.][]{ostriker2010}, this is not a concern. It is possible to fit slopes to the individual galaxy KS relation above $\Sigma_g > 10$ M$_{\odot}/$pc$^2$.  These slopes are 2.39, 2.36, 2.12, 2.14 and -2.27 for cases FUVFB50, FUVFB10, FB50, FB10 and FUV$^{\star}$, respectively.  \citet{kennicutt1998} notes that slopes from 1.29 up to 2.47 can be fit to an overall sample of galaxies in this range depending on the fitting approach employed. This calls into question the value of fitting slopes in a limited region near 10 M$_{\odot}/$pc$^2$. 

We do note that cases with higher feedback provide a modestly better fit but this is not particularly significant.  In particular, our chosen range of supernova energies is acceptable but the fit favours higher energies in this instance. Further, our simulations do not include the full range of early stellar feedback physics. It is possible that stellar winds, radiation pressure or photoionization play a role in setting the KS slope in high surface density regime, which could also account for the steeper slopes observed. 

We wish to be clear regarding
the interpretation of our plotted Kennicutt-Schmidt relation. Firstly, the goal is not to match the general relation for galaxy samples, such as that of \citet{kennicutt1998}, with its commonly cited power-law of $\Sigma^{1.4}$.  In the relevant region, near 10 M$_{\odot}$/pc$^2$ in the left panels of figure~\ref{fig:kshi}, the general relation not only steepens but the spread becomes so large that merely being inside the range of the data is fairly meaningless. Rather, we are comparing our specific galaxy to the individual relation for NGC 5055 which has similar radial profiles.  In this case, we see quite similar curves (as shown in the right panels of the figure) except for the FUV-only case.  A key point is how tight the relation is for a specific galaxy.

 Compared to NGC 5055,  the simulations have a similarly tight relation with a moderately steeper slope.  As outlined in Paper I,  high stellar densities imply a higher total pressure and more star formation. This connection has been demonstrated observationally by \citet{gallagher2018}, among others.   Referring back to Figure~\ref{fig:sd}, NGC 5055 has a systematically higher stellar surface density, $\Sigma_*$, at each radius but accurate scale heights are needed to convert that to a density.  We note that the similarity to NGC 5055 is somewhat coincidental. The \textsc{agora} galaxy was not intended to fit NGC 5055.  Several key properties (e.g. the rotation curve, giving the dark matter vertical gravity, and the stellar scale heights) were not matched.  Increasing the stellar scale height can dramatically reduce the gravity due to stars and lower the pressure and star formation rates.  Thus a galaxy model that better approximates NGC 5055 would be required before we could investigate the offset and slope in more detail.  Matching the high-surface density, innermost regions well might also require a model for molecular hydrogen which is expected to dominate there.

If we look to lower surface densities, $\Sigma_g \lesssim 3$ M$_{\odot}$/pc$^2$, the star formation rates appear consistent in all cases, including FUV$\star$.  We expect FUV heating to be most important in the outer regions of disk galaxies. Based on these first results, it appears that simulated galaxies require supernovae feedback for the inner-mid disk and FUV heating may be sufficient for the outer disk.  

Furthermore, FUV impacts the outer disk in a way that supernova feedback cannot. As the FUV mean free path is long, star formation in the inner regions of the disk can heat the outer regions of the disk. This effect is shown in Figure \ref{fig:truncation}, where the purple and green lines denote the star formation rate surface density for galaxies with SNe feedback only and SNe with FUV feedback, respectively. In the outer disk, beyond $\sim 6$ kpc, there is a change in slope for the case with FUV added: star formation is suppressed to a larger degree. This is something that SNe alone cannot accomplish: increasing the SNe energy merely changes the normalization of this relation, not the slope (see bottom panel Figure \ref{fig:truncation}).
 
 \begin{figure*}
 	\includegraphics[width=0.9\textwidth]{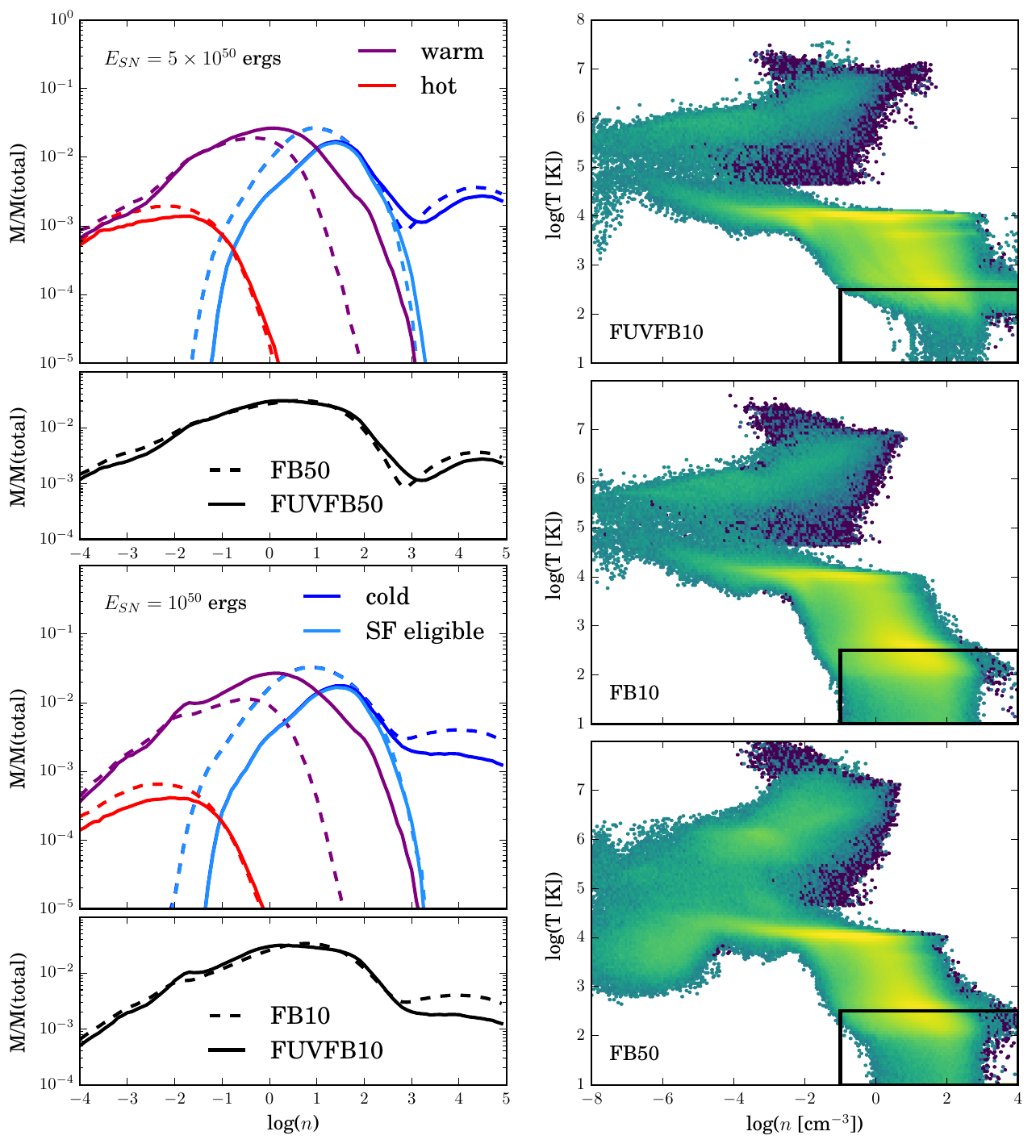}
	\caption{The density distribution for gas in different phases of the ISM. Hot gas (red) is any gas above $10^5$ K and this gas has most likely been heated by super-bubble feedback.  Warm gas (purple) is any gas between $10^5$ K and $2000$ K. Cold gas (blue) is any gas below $2000$ K.  Cold gas which is eligible for star formation (SF eligible, light blue) is any gas below 2000 K which has not been heated by super-bubble feedback. The black lines show the distribution for all of the gas. Dashed lines denote cases that have only super-bubble feedback, while solid lines have both super-bubble and FUV heating. Shown another way, the phase diagrams for simulations FB10noCR and FUVFB10noCR. The phase diagrams are weighted by mass, with lighter colours denoting higher mass. The inclusion of FUV introduces more warm diffuse gas, which is more in line with what we expect in nature.}
	\label{fig:allphase}
 \end{figure*}

\section{The phases of the ISM} \label{sec:phases}
In section \ref{ssec:pressure} we saw that adding adding additional feedback on top of supernovae feedback decreased their role in maintaining pressure support.  We are also interested in how the inclusion of self-consistent FUV heating impacts the structure of the ISM. We begin by considering how gas is divided into the different temperature phases of the ISM. We make divisions for these phases as follows. Hot gas is any gas above $10^5$ K and this gas has most likely been heated by super-bubble feedback.  Warm gas is any gas between $10^5$ K and $2000$ K. Cold gas is any gas below $2000$ K.  FUV is the dominant heating mechanism for any gas that has not been heated by supernovae. We add an additional category, cold gas which is eligible for star formation. This is any gas below 2000 K which has not been heated by super-bubble feedback. This gas can form stars if its density exceeds 100 cm$^{-3}$.

In the left column of Figure \ref{fig:allphase} the distribution of gas through different density bins is plotted for each of the temperature divisions outlined above. The top and bottom row are separated by the strength of the super-bubble feedback energy. The first noticeable difference is in the amount of warm gas: the inclusion of radiative transfer appreciably changes the distribution of warm neutral gas. The peak of the warm gas distribution remains near $n\sim0.3$ cm$^{-3}$ in all cases.  This matches expectations for the warm neutral phase of the ISM \citep{tielensBook}. However, when FUV heating is added on top of super-bubble feedback there is more warm gas at higher densities, between 10 and 10$^3$ cm$^{-3}$. This can be quantified by looking at the maximum density of the warm phase gas in each case. Considering the top two plots, adding FUV heating increases the maximum density of the warm phase gas by 1-2 orders of magnitude.

A complementary phenomenon is seen for the density distribution of cold gas.  There is a decrease in the amount of both cold diffuse and cold dense gas when the FUV heating is added (see cases FUVFB50 and FUVFB10).  We note that a  small fraction of the dense gas is not eligible for star formation because it is within two-phase particles. The portion of dense gas that is eligible for star formation is over-plotted as the light blue line in Figure \ref{fig:allphase}.  We make note of this mainly because using the super-bubble model maintains a trace of high density gas, $\gtrsim 10^3$ cm$^{-3}$, that is absent otherwise.   

These phase changes occur without significant changes to the total density distribution of the ISM. The black curves plotted in Figure~\ref{fig:allphase} show the total mass distribution at each density in the ISM for all of the temperature phases considered. The curves are very similar, with the only deviations being at the largest densities. This redistribution is particularly apparent when considering the warm medium. When FUV heating is included gas changes from being diffuse ($\lesssim 10$ cm$^{-3}$) and cold, to being diffuse and warm. However, the density structure at those densities remains mostly unchanged, only the temperature distribution is impacted. This can be seen very clearly in Figure \ref{fig:allphase}.

Another way to conceptualize this is by using the phase diagram, which plots the two-dimensional distribution of mass in the density-temperature space. In the right column of Figure \ref{fig:allphase} we plot the phase diagrams for three of the galaxies. If we contrast the top two plots, showing the phase diagrams for FUVFB10 and FB10, we can see that the greatest differences occur at low temperatures. The black box is drawn to aid the eye in comparing the diagrams. The case with FUV radiation has significantly less gas at low densities (< 10 cm$^{-3}$) and low temperatures (< 500K). Gas at these low densities should not populate the cold neutral medium, it too diffuse. The inclusion of the radiation allows it to be replaced by warm neutral counterparts. This is something that supernova feedback alone cannot do. In the bottom plot, the result for FB50 is shown. This has five times more supernova energy than the case above (FB10) and a similar phase diagram.

This outcome can be understood through pressure requirements. In the top-left and top-middle panels of Figure \ref{fig:pressure}, a comparison of the relative amount of the total pressure in a $P_{hot}$, SNe heated component, and $P_{th}$, warm thermal component are shown. When FUV radiation is included a larger amount of the total pressure support comes from the thermal component, decreasing the amount of support that needs to come from a SNe heated component. In the phase diagram, this translates to a larger amount of warm gas. When we do not use the FUV radiation, in general the ISM is quite cold across a large range of densities and supernovae are only capable of making this gas hot or stirring it up. FUV provides us with an intermediate step, where we can have a high pressure warm medium, as seen in Figure \ref{fig:pressure}.  This shows a crucial role for FUV radiation in maintaining a phase of warm neutral gas.

\section{Summary \& Conclusions}
For the first time we have simulated a full, isolated galaxy using a self-consistent treatment for FUV   photo-electric heating via dust grains of the typical ISM, implemented through a ray-tracing approach to radiative transfer with absorption.   We used a relatively low dust opacity of $300$ cm$^2$/g (scaled linearly from solar metallicity)
as in \cite{wolfire2003} and a fixed heating efficiency of 3\%.  This can be taken as an upper bound on the effectiveness of this heating for the typical ISM.  We do not model molecular hydrogen and thus the dissociating effects of FUV are not included.
We have explored the effect of FUV alone and in combination with a highly-constrained model for supernova feedback.  

The Kennicutt-Schmidt relation for individual galaxies is quite tight compared to the overall relation covering the full population, which has spread of 1-2 dex.  The \textsc{agora} isolated galaxy we simulate here is fortuitously similar to NGC 5055.  We  get a similarly tight Kennicutt-Schmidt relation from simulations using both FUV and SNe together, including its small spread of $<$ 0.5 dex.    A small spread and significant slope variations are strongly indicated for individual galaxies \citep{ostriker2010} and thus  should be expected in simulations.  This result strongly supports the importance of secondary parameters, such as the stellar density, in setting the star formation rate, as championed by \citet{blitz2006}.  This shows that our full model is both physical and consistent with established relations.  The only free parameter is the energy per supernova and a factor of five variation still results in simulations that are reasonable fits.  The Kennicutt-Schmidt relation fit marginally favours higher net supernovae energy ($50 \%$ internal losses or $\sim 5\times10^{50}$ ergs per SN) and there are indications of a small slope difference.  These differences are far smaller than the spread in the full KS galaxies data set.  Better galaxy models will be required to confirm  these differences given that the \textsc{agora} galaxy model was not built to match NGC 5055 specifically. 

We find that FUV heating on its own is not sufficient to regulate star formation.  In this case, the galaxy rapidly consumes the available fuel.  This is in contrast to the \cite{ostriker2010} framework, which linked regulation to FUV heating.  The mode of failure of FUV regulation is important for future analytical attempts to model galactic star formation regulation.  Simulations with ineffective feedback result in a very clumpy ISM and this behaviour is qualitatively the same as no feedback at all. The driver of clump formation is converging flows in the $r-\phi$ plane which are implicitly excluded by assumed smooth radial density profiles in analytical models.  The moderate temperatures generated by FUV cannot unbind these large clumps and they dominate the evolution going forward, sweeping up most of the gas mass.  This behaviour was also seen in the \textsc{agora} isolated simulations of this galaxy \citep{agoraIC}.  These used either no feedback or purely thermal supernova feedback, which is well-known to be ineffective \citep{katz1992}.  The key difference from this work is that in \textsc{agora}, the star formation efficiency was very low (about 1\% per free-fall time) and a stiff equation of state prevented dense gas with short free-fall times.  This brought the total star formation rates closer to Milky Way-like expectations.  However, most of the mass is still in massive clumps exceeding 500 M$_{\odot}$/pc$^2$, which in turn produce large star clusters.

In combination with the supernova feedback, we see that the addition of FUV radiation has important impacts on the ISM. Our results confirm that FUV heating is non-local and that the gas state depends on distant sources ($\gtrsim$ 6 kpc) and correct absorption by the intervening material.  The inclusion of FUV heating helps regulate star formation in a more gentle way. FUV heating is able to produce regulated galaxies with realistic scale heights. We can see this when looking at the distribution of gas among the hot, warm and cold phases of the ISM.  FUV heating is able to move substantial mass from the cold gas into the warm neutral phase.  This provides significant pressure so that less star formation is needed, particularly in the outer disk.  Our analysis confirms that the required pressure (as employed in paper I) provides a robust framework to understand how different feedback mechanisms combine to regulate star formation and structure in the ISM of galaxies.  

A potential focus for future work is the outer disk where FUV provides a large portion of the support.  In our results, the drop off in star formation is significantly different for supernovae feedback runs with and without self-consistent FUV (Figure~\ref{fig:truncation}).   These differences may depend on how star formation is modeled.  In future work, we will explore more complex star formation models more closely tied to the cold, dense phase \cite[e.g.][]{semenov2018}.  Observationally, this low surface density regime is difficult to measure accurately but it could provide important insights into how star formation and ISM structure are linked to FUV, turbulence and other feedback.

This work indicates that fitting star formation expectations while producing a realistic ISM can provide powerful constraints on how different feedback processes function in real galaxies.  As in paper I \citep{benincasa2016}, we find that no one star formation or ISM diagnostic on its own can tell the whole story. The Kennicutt-Schmidt relation (Figure \ref{fig:kshi}) provides an illustration of this point. If we study the right panels of this figure, which feature direct comparisons to NGC 5055, all of the galaxies that were able to regulate their star formation provide a convincing match. We know from our other diagnostics that they have significantly different overall star formation rates, different scale heights and different ISMs.  Thus we should perform careful comparisons to observed galaxies involving all of these properties and not just star formation rates. In particular, the low supernova energy plus FUV case provides the best match to the scale height observations of \cite{patra2019} for NGC 5055 (Figure~\ref{fig:sfrH}).  Our galaxy is a passable proxy for NGC 5055 but there are differences in the profiles of both gas and stars, as well as the rotation curve.  In order to make these comparison properly, we should use galaxies that are designed specifically to match observed galaxies. This will be a target of future work.

\section*{Acknowledgements}
The authors thank the scientific editor, Joop Schaye, and the anonymous referee whose insight greatly improved the work presented here. The authors thank Charlotte Christiansen, Sarah Loebman and Andrew Wetzel for helpful discussions that improved the content of the manuscript. SMB acknowledges support from the Vanier Canada Graduate Scholarship program. JW and HMPC acknowledge support from NSERC. ARP acknowledges the support of The Japanese Society for the Promotion of Science (JSPS) KAKENHI grant for Early Career Scientists (20K14456). BWK acknowledges funding support in the form of a Postdoctoral Research Fellowship from the Alexander von Humboldt Siftung. Computations were performed on the Niagara supercomputer at the SciNet HPC Consortium. SciNet is funded by: the Canada Foundation for Innovation; the Government of Ontario; Ontario Research Fund - Research Excellence; and the University of Toronto. This work was made possible by the facilities of the Shared Hierarchical 
Academic Research Computing Network (SHARCNET:www.sharcnet.ca) and Compute/Calcul Canada. This work was made possible by the facilities of the Shared Hierarchical Academic Research Computing Network (SHARCNET:www.sharcnet.ca) and Compute/Calcul Canada.

\section{Data Availability}
The data underlying this article will be shared on reasonable request to the corresponding author.



\bibliographystyle{mnras}
\bibliography{paper3} 




\appendix

\section{Radiative Transfer Comparisons}

\begin{figure}
	\includegraphics[width=0.45\textwidth]{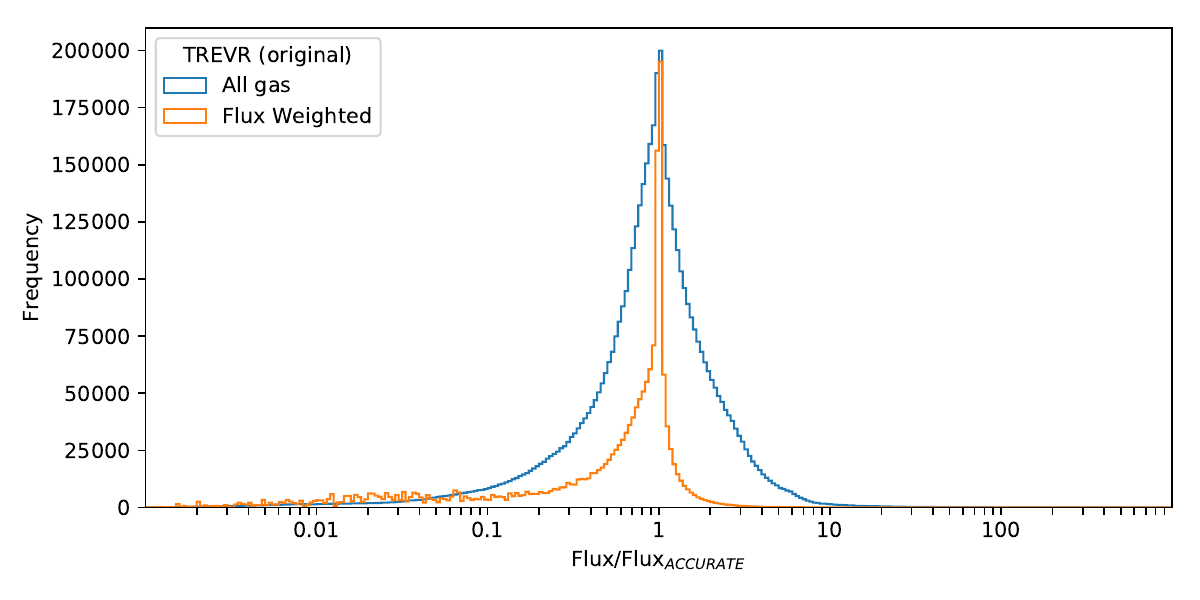}
	\caption{Flux errors for original TREVR (used in the current work) compared against an extremely accurate flux estimate using the most recent version ($\tau_{\rm REFINE}=0.01$).  The all gas curve is mass weighted and includes regions receiving low or negligible flux.  The flux-weighted distribution shows errors weighted by the total flux received.}
	\label{fig:fluxerror}
\end{figure}

\begin{figure}
	\includegraphics[width=0.45\textwidth]{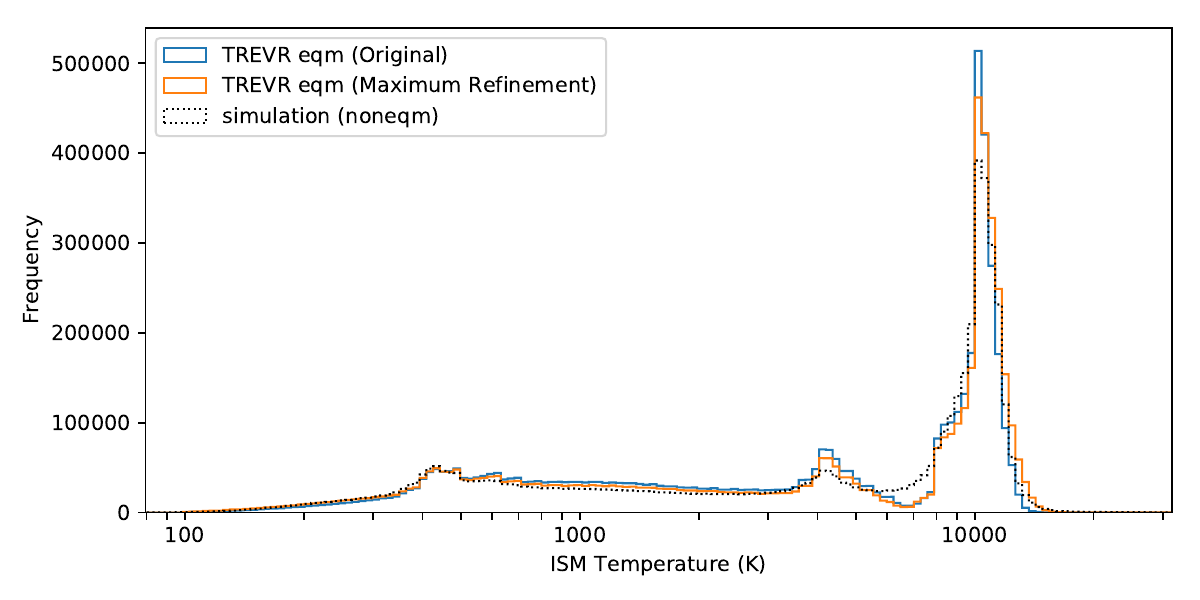}
	\caption{Equilibrium temperature distribution (including just the dominant photoelectric heating and radiative cooling) for original TREVR (used in the current work) compared to an extremely accurate flux estimate using the most recent version ($\tau_{\rm REFINE}=0.01$). The dotted curve is the non-equilibrium temperature distribution in the full simulation output that was used.}
	\label{fig:tempdist}
\end{figure}

This work employed the original version of TREVR \citep{woodsPHD}.  The most recent version, as discussed in \citet{grond2019}, has a modified ray trace and adaptive ray refinement to control errors.  It does this by estimating the maximum optical depth error for each ray segment.  If this exceeds the $\tau_{\rm REFINE}$ parameter, the ray is split and re-tested until it succeeds or reaches the resolution limit.  \cite{grond2019} showed that this typically limits the overall optical depth errors to $\tau_{\rm REFINE}$/10 (with similar flux errors if $\tau_{\rm REFINE} \lesssim 1$).  Thus for $\tau_{\rm REFINE}=0.01$ the results are equivalent to full resolution ray traces which provides an accurate answer for comparisons.  Conversely, $\tau_{\rm REFINE}=\infty$ means no added refinement for the new code and results in similar levels of error (but not the same results) as the original code.  The new code is considerably slower (by more than an order of magnitude) than the original version used here and thus impractical to use for full runs.  However, we applied it on a single evolved output at 400 Myr (FUVFB10) where star formation rates and the ISM properties have reached a statistical steady state, allowing us to estimate the accuracy of the radiative transfer used for the current work.

Figure~\ref{fig:fluxerror} shows the flux errors for the representative output using the original scheme relative to an extremely accurate estimate (TREVR with $\tau_{\rm REFINE}=0.01$).  The wider distribution is for all gas, including gas with relatively little flux (i.e. where FUV is a minor contributor to the heating).   The narrow distribution is flux weighted and demonstrates TREVR's small, absolute flux errors. The majority of the gas is within a few percent of the accurate prediction.   This avoids significant differences in the FUV heating.   We characterize this via a temperature histogram for the ISM, shown in figure~\ref{fig:tempdist}.   As we were unable to perform a full run with the extremely accurate version, we computed equilibrium temperatures with just photoelectric heating and radiative cooling for both the original code and accurate error controls on a single output.  We focus on ISM conditions, where FUV is the dominant heating.  The distributions are nearly identical.  For comparison, we include the non-equilibrium distribution from the raw simulation output (dotted), which is almost the same.  This demonstrates the dominance of FUV heating in this regime.  

\begin{figure}
	\includegraphics[width=0.47\textwidth]{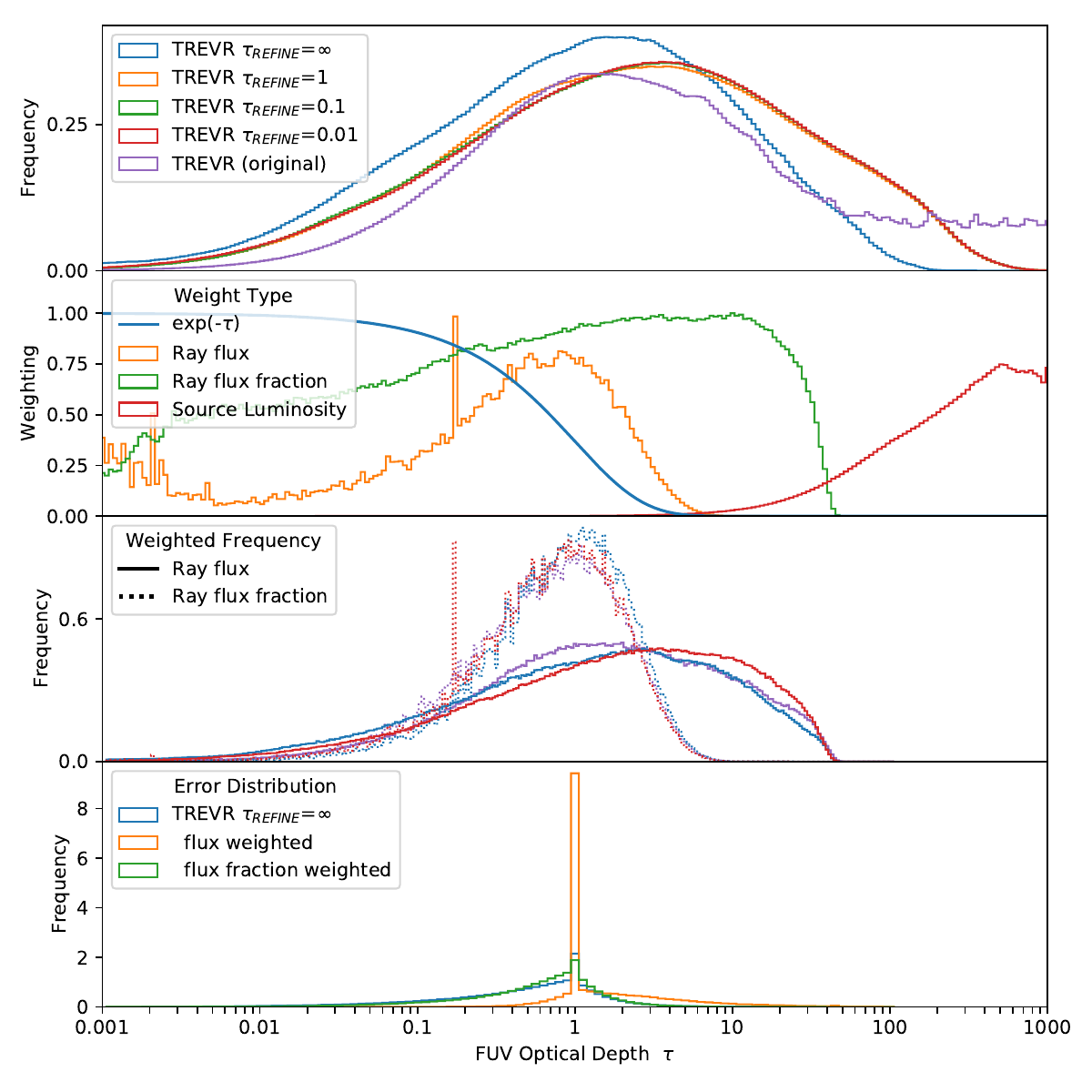}
	\caption{Detailed ray-by-ray comparison of the original TREVR to the most recent version with different refinement criteria.  The original TREVR had no refinement error control, as does the newest version when $\tau_{\rm REFINE}=\infty$.  The new version becomes extremely accurate for $\tau_{\rm REFINE} \lesssim 0.1$.  The top panel shows the frequency of FUV optical depths per ray.  The second panel shows various ways to weight the importance of they rays: attenuation ($e^{-\tau}$),  received flux by the gas from the ray, received flux as a fraction of the total received by that particle and source luminosity.  This confirms that most luminous sources are blocked by high columns and contribute little.  The third panel shows the optical depth weighted for each ray's flux and flux fraction.  The bottom panel shows the per ray optical depths for the least accurate TREVR relative to the most accurate TREVR refinement tested ($\tau_{\rm REFINE}=0.01$).}
	\label{fig:taudist}
\end{figure}

Radiative transfer with absorption is very challenging numerically, especially when a single resolution element can be optically thick, such as for UV in galaxies.   Small offsets can dramatically change received fluxes.  Following the appendix in \citet{hopkins2012}, we examined the optical depth distribution between sources and gas and show the results in figure~\ref{fig:taudist}.   We focus on the importance of examining the optical depth errors for radiation bands of interest, rather than the full column density distribution.  For FUV, columns less than $10^{21}$ cm$^{-2}$ are most important, as this is where the optical depth is $\lesssim 30$ (this may be compared to figure A1 in \citet{hopkins2012} for the local absorption estimator, LEBRON).  

The top panel of figure~\ref{fig:taudist} shows the raw distribution of optical depths for TREVR on the representative output with different accuracy criteria applied. One particle in 100 was used and each receives input from $\sim 1400$ rays.  The differences are largest for extreme optical depths ($\gtrsim 30$).  To extract useful information from this distribution we must weight the rays based on their effect.  The second panel shows potential weights: attenuation ($e^{-\tau}$), flux at the receiving gas and the relative flux contribution (out of all rays hitting the specific gas particle).  The second weighting emphasizes rays that deliver high fluxes.  The third weighting selects rays that are most important for a specific gas element (i.e. influencing its heating and temperature).  We also show luminosity weighting.  This last is less useful as it favours rays from sources that are highly obscured and contribute little to the radiation field for the vast majority of the gas.   When we convolve the distributions with weights that reflect the impact on the gas, we see that the original TREVR is fairly similar to the new version.   The bottom panel shows a distribution of optical depth ratios, where $1$ indicates a match to the most accurate TREVR run.  These distributions, tightly clustered around the correct optical depth, enable the good overall flux error distribution shown in figure~\ref{fig:fluxerror}.

Naturally, the most intensely irradiated gas is within a few pc of bright, young sources.  However, these sources tend to be deeply embedded. The vast majority of the ISM experiences moderate fluxes.  The third panel of figure~\ref{fig:taudist} shows that rays with optical depths between 0.1 and 10 are most important in determining the flux in the typical ISM, corresponding to sources that are either more evolved and/or have more favourable sight-lines.  Half the flux for a typical parcel of gas is from distances of 6.4 kpc or more.  These factors confirm the importance of absorbing material in the several kpc between the gas and its influential sources.

\bsp	
\label{lastpage}
\end{document}